\documentclass{article}

\usepackage{arxiv}

\usepackage[utf8]{inputenc} 
\usepackage[T1]{fontenc}    
\usepackage[hidelinks, hyperfootnotes=false]{hyperref}       
\usepackage{url}            
\usepackage{booktabs}       
\usepackage{amsfonts}       
\usepackage{nicefrac}       
\usepackage{microtype}      
\usepackage{lipsum}         
\usepackage{doi}

\usepackage[numbers]{natbib}
\usepackage{graphicx}

\usepackage{here}
\usepackage{color}
\usepackage{multirow}

\title{Massive geolocation data reveal evacuation behaviour during the 2024 Noto Peninsula earthquake and tsunami}

\date{}

\author{ \href{https://orcid.org/0000-0001-9247-4104}{\includegraphics[scale=0.06]{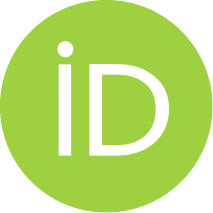}\hspace{1mm}Fumiyasu~Makinoshima}\thanks{Corresponding author: Fumiyasu Makinoshima (fumiyasu.makinoshima.a5@tohoku.ac.jp)} \\
	International Research Institute of Disaster Science\\
	Tohoku University\\
	Sendai, Japan \\
	\texttt{fumiyasu.makinoshima.a5@tohoku.ac.jp} \\
	\And
	\href{https://orcid.org/0000-0002-9920-5107}{\includegraphics[scale=0.06]{orcid.pdf}\hspace{1mm}Saki~Yotsui} \\
	Research Center for Advanced Science and Technology\\
	The University of Tokyo\\
	Tokyo, Japan\\
	\texttt{yotsui@city.t.u-tokyo.ac.jp} \\
    \And
    \href{https://orcid.org/0000-0001-7585-2046}{\includegraphics[scale=0.06]{orcid.pdf}\hspace{1mm}Shosuke~Sato}\\
	International Research Institute of Disaster Science\\
	Tohoku University\\
	Sendai, Japan \\
	\texttt{shosuke.sato.a7@tohoku.ac.jp} \\
    \And
    \href{https://orcid.org/0000-0001-7628-575X}{\includegraphics[scale=0.06]{orcid.pdf}\hspace{1mm}Fumihiko~Imamura}\\
	International Research Institute of Disaster Science\\
	Tohoku University\\
	Sendai, Japan \\
	\texttt{fumihiko.imamura.c3@tohoku.ac.jp} \\
}

\renewcommand{\shorttitle}{\textit{arXiv} Template}

\hypersetup{
pdftitle={Massive geolocation data reveal evacuation behaviour during the 2024 Noto Peninsula earthquake and tsunami},
pdfsubject={physics.soc-ph},
pdfauthor={F. Makinoshima et al.},
}

\begin{document}

\maketitle

\begin{abstract}
On 1 January 2024, devastating tsunamis caused by the Noto Peninsula earthquake hit coastal areas within several minutes, but only two tsunami casualties were officially reported.
Despite its importance, the cause of this unexpectedly low human loss was unclear because of the limited access to the peninsula and the presence of many visitors during the holiday, which made conducting conventional surveys infeasible.
Here, we reveal evacuation behaviour during the 2024 Noto Peninsula tsunami using massive geolocation data collected from a smartphone app.
By analysing these massive data, which include over 1.5 million records collected on this day, we find that the evacuation was extremely fast, occurring within 2--6 minutes after the origin time.
Further analyses suggest that these fast departures were driven mainly by strong ground shaking; the fact that the tsunami occurred during the family-oriented New Year holiday was also a key factor.
Additionally, the long-term analysis of the data reveals that people started returning to the coastal area 20--100 minutes after the origin time, which was long before the downgrading and cancellation of the tsunami warnings.
These results highlight the utility of the innovative data-driven approach to evacuation surveys, which addresses the limitations of conventional evacuation surveys.
\end{abstract}

\section*{Introduction}
\footnote{This work has been published in {\it Communications Earth \& Environment}. The final version is available~\cite{makinoshima_2025}.}
On 1 January 2024, at 16:10 JST (7:10 UTC), the Noto Peninsula earthquake, of magnitude M$_{\mathrm{w}}$ 7.5, occurred and generated a large tsunami~\cite{fujii_2024,kutschera_2024}.
The coastal areas facing the Sea of Japan were hit by strong ground motion, and large tsunamis caused devastating inundations that completely destroyed buildings in some areas~\cite{yuhi_2024a}.
Numerical tsunami simulations suggest that the earthquake not only directly caused tsunamis but also caused submarine landslides that became additional tsunami sources~\cite{mulia_2024}.
Owing to the multiple tsunami sources and an earthquake epicentre quite close to the shore, tsunamis arrived in areas close to the Noto Peninsula within several minutes after the earthquake~\cite{masuda_2024}.
Additionally, tsunamis continued for a long time because of the diffractions and reflections caused by complex geographical features around the peninsula~\cite{yuhi_2024b}, which likely caused delayed excitations of large tsunami amplitudes.
Consequently, people in coastal areas had limited lead time for evacuation during this tsunami event and had to remain in safe places after evacuating.

Previous tsunami evacuation survey results suggest that in tsunami disasters, people typically take some time to begin evacuating rather than performing immediate evacuation.
This delayed evacuation can result in tremendous human loss, as was seen in past tsunami disasters such as the 2004 Indian Ocean tsunami and the 2011 Tohoku tsunami, which caused more than 230,000 and 22,000 casualties, respectively~\cite{mori_2022}.
For example, during the 2011 Tohoku tsunami, local residents took approximately 14--34 minutes to begin evacuation after the earthquake~\cite{makinoshima_2024}.
When people face tsunami risk, they tend to take some actions before evacuation, i.e., milling~\cite{wood_2018}.
Existing surveys of past tsunami-related events reported that such actions include collecting information, preparing for evacuation, confirming safety, and gathering family members~\cite{lindell_2015,fraser_2016,blake_2018,makinoshima_2020}.
A detailed fact-finding survey for the 2011 Tohoku event revealed that people performed exceedingly complex evacuation processes involving these actions, even when they had a sufficient understanding of tsunami risk~\cite{makinoshima_2021}.
Thus, it is quite challenging for people to perform immediate tsunami evacuation behaviours, resulting in several tens of minutes passing before evacuation departure.

Although the 2011 Tohoku tsunami caused severe human casualties of over 22,000, a statistical comparison between the 2004 Indian Ocean tsunami and the 2011 Tohoku tsunami revealed many fewer tsunami casualties during the 2011 event, even with similar tsunami heights and arrival times~\cite{suppasri_2016}; it was suggested that the evacuation was generally performed before the tsunami arrival time (30 to 60 minutes~\cite{muhari_2012}), contributing to drastically reduced tsunami casualties.
Such successful evacuation trends, similar to those observed during the 2011 Tohoku tsunami, were, however, not sufficient to reduce the number of tsunami casualties during the 2024 Noto Peninsula tsunami event since the tsunami arrived extremely quickly.
Accordingly, considerable human loss was initially expected for this event on the basis of statistics from the past mega-tsunamis.
However, surprisingly, only two tsunami casualties among the total casualties of more than 200 people were officially reported in this event, which was much lower than the number estimated from the extremely fast tsunami arrival and observed tsunami heights.

Clarifying the factors resulting in this surprisingly small amount of human loss would provide essential knowledge for reducing human loss during future mega-tsunamis caused by large earthquakes such as the Cascadia megathrust earthquake~\cite{satake_1996} and the Nankai megathrust earthquake~\cite{ando_1975}, in both of which fast tsunami arrival is expected~\cite{wood_2012,hirano_2023}.
However, since the limited access routes to the peninsula were greatly damaged by strong ground shaking and even emergency rescuers experienced difficulty accessing these areas~\cite{egawa_2024}, fact-finding surveys on evacuation behaviours could not be conducted immediately after the event.
In addition, because the earthquake and tsunami occurred during the Japanese New Year holiday, the survey subjects, most of whom were visitors, left the area soon after access to the peninsula recovered.
Under such circumstances, the conventional evacuation survey approach involving questionnaires~\cite{gaillard_2008,fraser_2016,blake_2018} or interviews~\cite{dudley_2011,makinoshima_2021}, which typically takes time to prepare and conduct, could not be used in this event.
Consequently, despite its importance, the overall picture of evacuation behaviour during this event that resulted in a small number of tsunami casualties was unclear.

\begin{figure}[H]
	\centering
	\includegraphics[width=0.85\textwidth]{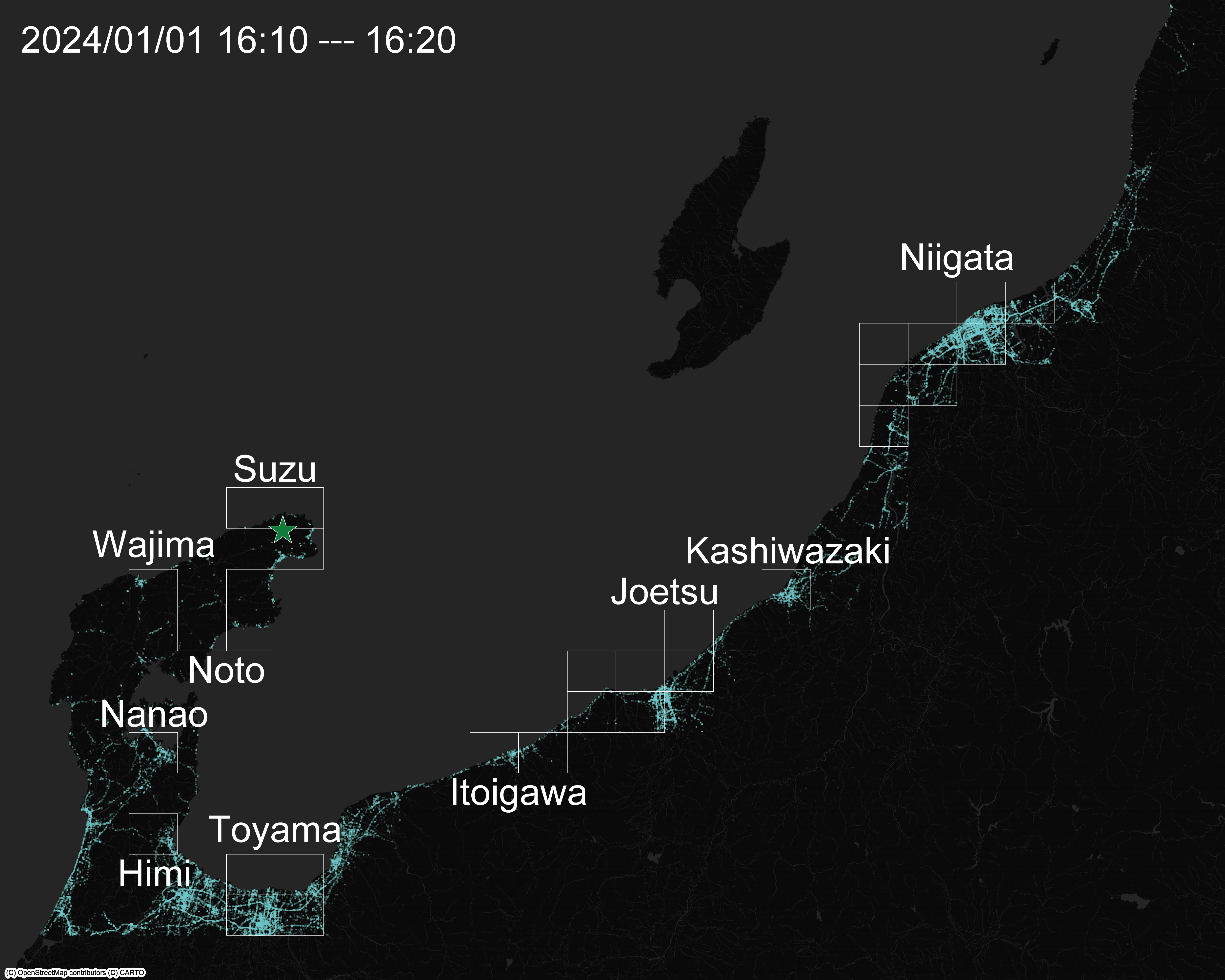}
	\caption{Massive geolocation data recorded during the 2024 Noto Peninsula earthquake and tsunami (16:10--16:20 JST).
    An animation is available on Zenodo (\url{https://doi.org/10.5281/zenodo.14300988})~\cite{zenodo}.
    Each blue dot represents an individual position recorded at a certain time.
    The animation reveals notable landward movements after the earthquake occurred.
    In this study, 10 coastal cities (Wajima, Suzu, Noto, Nanao, Himi, Toyama, Itoigawa, Joetsu, Kashiwazaki, and Niigata) with sufficient observations were analysed.
    The areas for the analyses were determined on the basis of the standard geological grid mesh code (JIS X 0410), shown as white rectangles in the figure.
    The green star represents the epicentre of the main shock.}
	\label{fig:fig1}
\end{figure}

Here, we find that massive geolocation data collected from a smartphone app recorded the evacuation behaviour during the 2024 Noto Peninsula earthquake and tsunami (Fig.~\ref{fig:fig1} and Supplementary Movie~\cite{zenodo}).
By analysing this massive geolocation data, including over 1.5 million records, this study reveals extensive details of the evacuation behaviour during the tsunami event in a broad area, spanning over three prefectures, i.e., Ishikawa, Toyama, and Niigata, where tsunami warnings were delivered.
The analysed data reveal several important aspects of tsunami evacuation, such as evacuation departure time, evacuation triggers, and successive long-term behaviour under warning updates.
This study not only provides valuable insights into tsunami preparedness efforts for future mega-tsunamis but also demonstrates a novel data-driven approach to evacuation studies, which addresses the limitations of conventional evacuation surveys, such as memory errors, cognitive bias, and mental burden on survivors.

\section*{Results}
\subsection*{Evacuation departure timing}
\begin{figure}[H]
	\centering
	\includegraphics[width=0.85\textwidth]{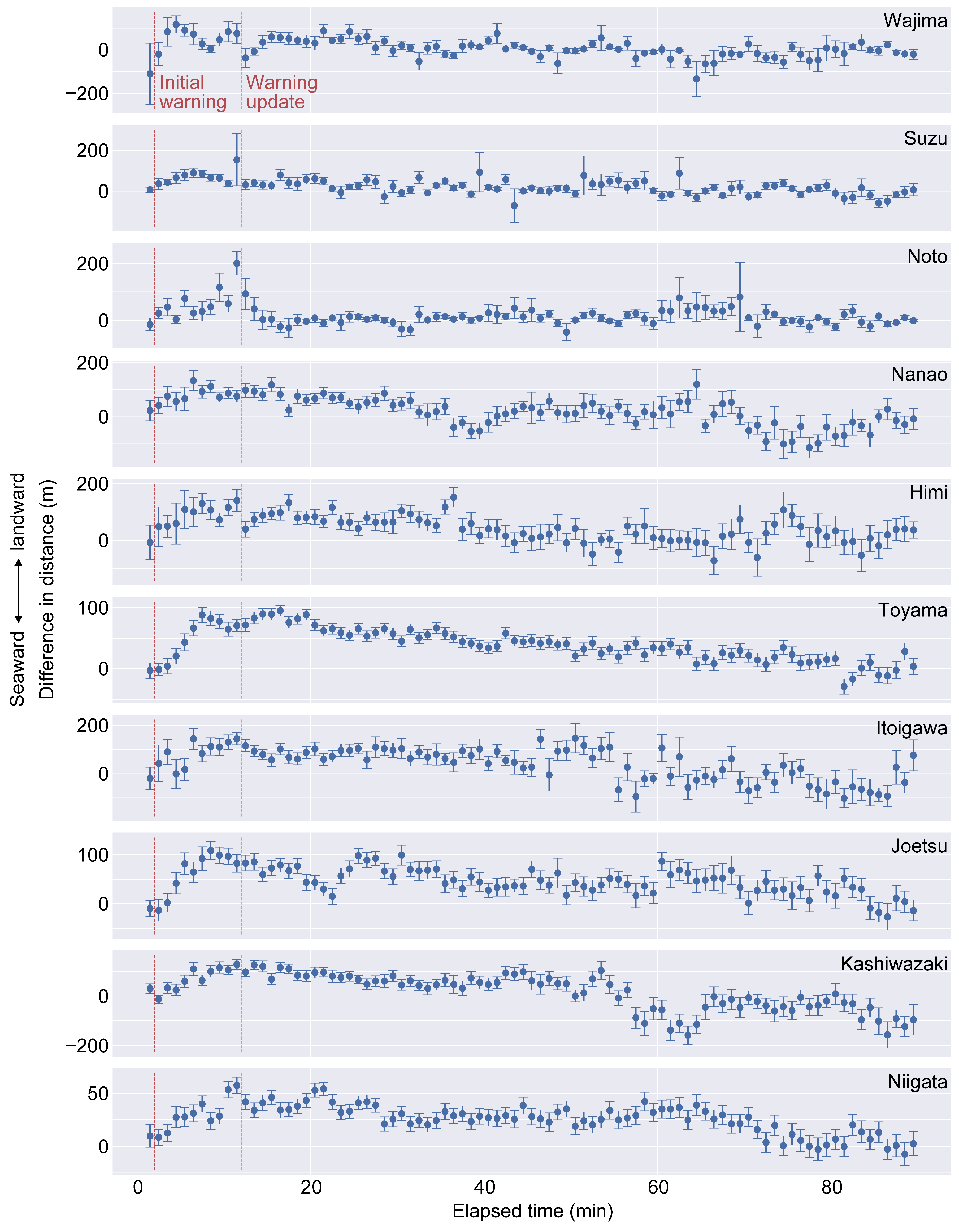}
	\caption{Evacuation departure trends.
    The difference in distance between each location data point and the coastline were calculated, and the differences were organised into 1~min bins.
    The plots and bars represent the mean values and standard errors, respectively.
    Changes in the positive direction indicate landward movements, and those in the negative direction indicate seaward movements.
    The increasing trends of landward movement within 10 minutes after the origin time are considered to represent tsunami evacuation behaviour.}
	\label{fig:fig2}
\end{figure}

To investigate the evacuation departure timing, we analysed the geolocation data recorded within 90~min of the occurrence of the earthquake (Fig.~\ref{fig:fig2}).
We calculated the difference in distance between each location data point and the coastline, and the data were organised into 1~min bins to show the trends (see Methods for details).
The results revealed that in cities on the Noto Peninsula (Wajima, Suzu, Noto, Nanao, and Himi) and in Itoigawa, the apparent increasing trend of landward movements started 2--3 minutes after the earthquake, indicating fast evacuation departures immediately after the first warning from the Japan Meteorological Agency (JMA), i.e., 2~minutes after the earthquake.
In contrast, in the other areas (Toyama, Joetsu, Kashiwazaki, and Niigata), the apparent increasing trend of landward movements started after a short delay, i.e., 4--6 minutes after the earthquake.
However, these evacuation departure trends occurred before the major tsunami warning, an update of the initial tsunami warning 12 minutes after the origin time, was issued.
In all the areas analysed in this study, the increasing trends of landward movements generally ended within 10 minutes after the origin time, suggesting that evacuation was completed within a short period.
These results suggest that people generally initiated evacuation behaviour within several minutes and arrived at safe places within a short period during this tsunami event.

\subsection*{Evacuation triggers}
\begin{figure}[H]
	\centering
	\includegraphics[width=0.425\textwidth]{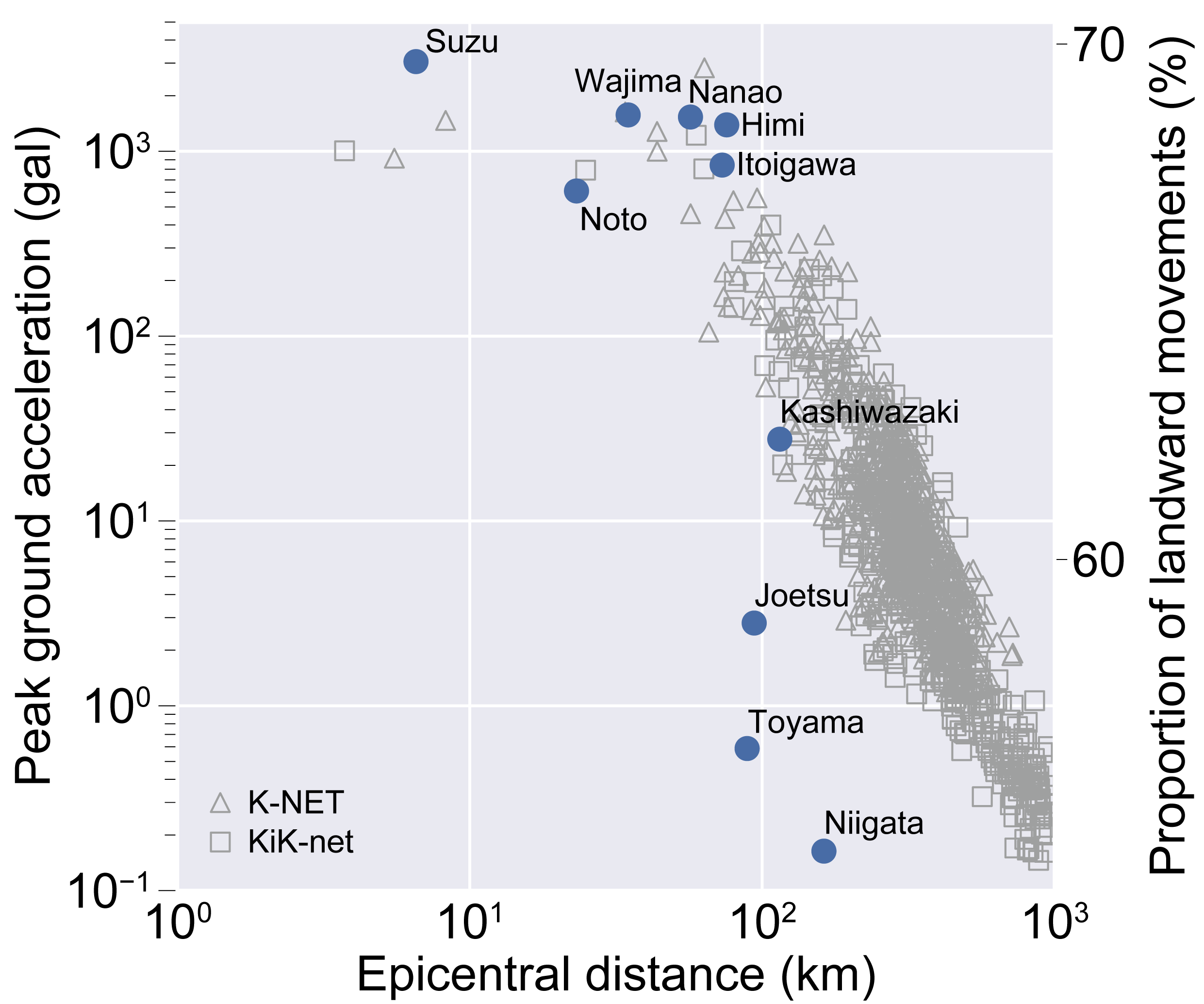}
	\caption{Relation between the observed peak ground accelerations and the proportion of landward movements within 10 minutes after the earthquake origin time.
    Locations of municipal offices were used to calculate the distance from each city to the epicentre.}
	\label{fig:fig3}
\end{figure}

We analysed the proportion of landward trips within 10 minutes of the origin time as a measure of the intensity of evacuation activities and found that they could be roughly divided into two groups, i.e., higher-intensity cities (Wajima: 68.5\%; Suzu: 69.6\%; Noto: 67.0\%; Nanao: 68.5\%, Himi: 68.3\%; Itoigawa: 67.5\%) and lower-intensity cities (Toyama: 56.7\%; Joetsu: 58.9\%; Kashiwazaki: 62.2\%; Niigata: 55.0\%).
The group with higher evacuation intensity contains cities in different prefectures (Ishikawa, Toyama, and Niigata), suggesting that this higher intensity was not due to the level of disaster preparedness or level of information transmission in different prefectures.
We found that this difference in evacuation intensity was related to the intensity of ground shaking (Fig.~\ref{fig:fig3}).
The ground acceleration observed during this event~\cite{nied} rapidly decreased at a distance from the epicentre of approximately 100~km, and the observed difference in evacuation intensity corresponded to this decay in ground acceleration.
This result indicates that the primary cause of the higher evacuation intensities was strong ground motions.
The cities with higher evacuation intensities were those in which faster evacuation departures (2--3 minutes from the origin time) were observed.
This consistency indicates that the primary evacuation trigger in these areas was strong ground motion rather than warning issuance, considering the very short time interval of 1--2 minutes between the first warning issuance and the increasing trend of landward movements.
Additionally, this result indicates that the tsunami warning was the primary evacuation trigger in cities where slightly delayed evacuation departure trends were observed.

\subsection*{Return from evacuation sites}
\begin{figure}[H]
	\centering
	\includegraphics[width=0.85\textwidth]{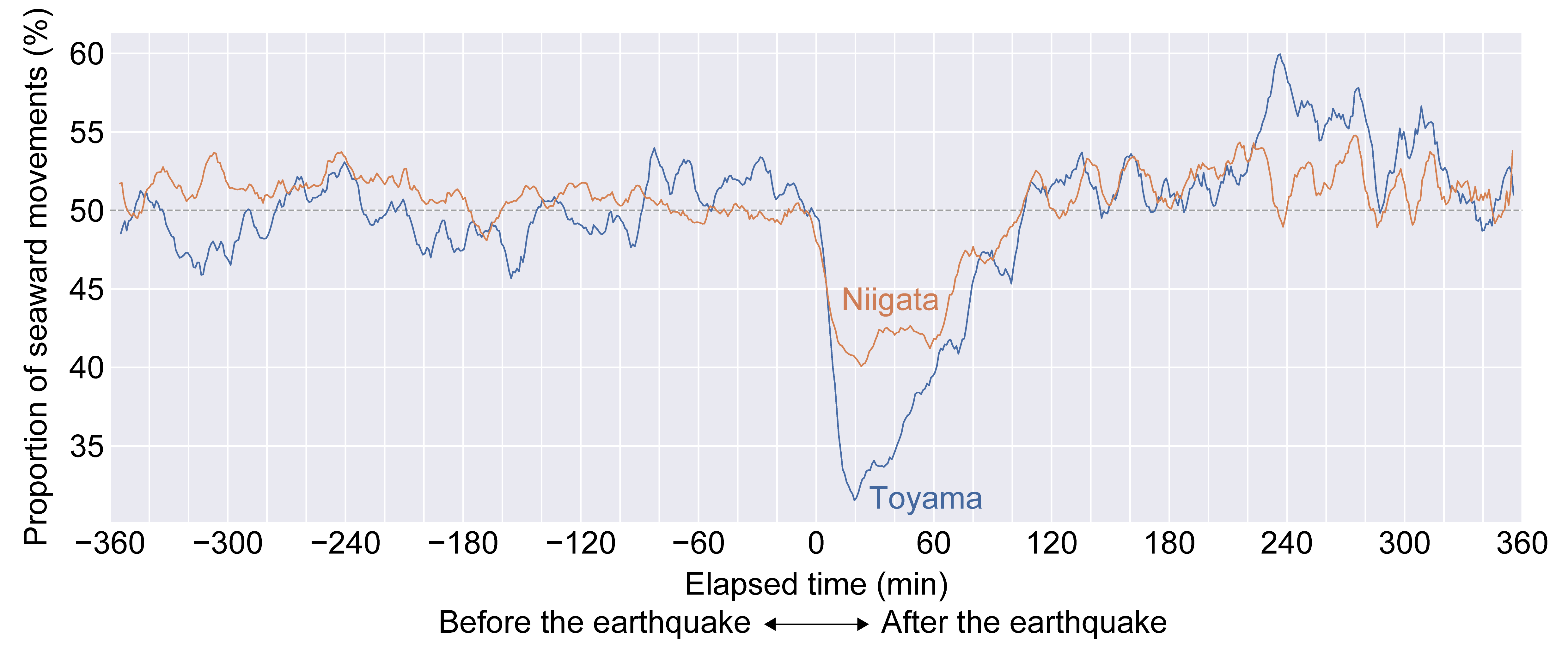}
	\caption{Time series of the proportion of seaward trips within 6 hours before and after the earthquake origin time.
    Here, moving averages with a 10 minute window size are presented to show the trends.}
	\label{fig:fig4}
\end{figure}

We extended the time range of the analysis to 6 hours before and after the origin time to examine long-term behaviour after the emergency evacuation (Fig.~\ref{fig:fig4}).
We focused on behaviours in Toyama and Niigata since these areas did not experience severe damage; thus, people could return to coastal areas, and these areas had larger sample sizes.
In both areas, the proportion of seaward movements consistently decreased over 20 minutes after the earthquake occurred.
Although different rates of increase were observed in different areas, 20 minutes after the earthquake occurred, the trend of seaward movements became positive.
This positive trend continued for several tens of minutes, and the trend of movements then returned to almost normal ($\sim 50\%$) 100 minutes after the origin time.
These results suggest that in both areas, people stopped evacuating and returned to coastal areas from inland evacuation sites 20--100 minutes after the earthquake occurred.
During this tsunami event, the major tsunami warning was downgraded to a tsunami warning at 20:30 on 1 January (JST), and the tsunami warning was later downgraded to a tsunami advisory at 1:15 on 2 January (JST).
The tsunami advisory was finally lifted at 10:00 on 2 January.
Therefore, the results of this study suggest that people subjectively evaluated the tsunami risk and returned to coastal areas before a low tsunami risk was officially determined.
On the basis of previous tsunami events that occurred in the Sea of Japan, JMA expected that large tsunamis could continue due to diffractions and reflections and carefully updated the tsunami warning according to observations.
In fact, the maximum tsunami amplitude observed at Kanazawa in Ishikawa prefecture occurred at 19:09 on 1 January (JST).
Thus, the time of return to coastal areas estimated from the analysis was much earlier than the arrival time of the maximum tsunami amplitude.
Overall, while people exhibited very fast evacuation departure trends, they also stopped evacuating within a short time during this tsunami event.

\section*{Discussion}
In this study, we examined massive geolocation data collected via a smartphone app that recorded evacuation behaviour during the 2024 Noto Peninsula earthquake and tsunami.
By analysing the geolocation data, we extensively explored the evacuation behaviours in three prefectures, which had been unclear due to the difficulty of accessing damaged areas and the presence of many visitors during the New Year holiday.
Although few studies have used geolocation data to study tsunami evacuation, one study~\cite{togawa_2018} aimed to reveal tsunami evacuation behaviour via aggregated geolocation data.
Since existing aggregate data have a coarse spatiotemporal resolution of 250~m every hour, determining tsunami evacuation dynamics via coarse spatiotemporal geolocation data has been difficult.
In contrast, this study used a disaggregated geolocation data sample with high frequency (1~min at maximum) and demonstrated that tsunami evacuation behaviours can be determined with such data.
This novel approach for evacuation surveys using massive and fine-grained geolocation data can reveal evacuation behaviours in detail while addressing the limitations of conventional surveys that use questionnaires and interviews, i.e., memory errors, psychological bias, and mental burden on subjects~\cite{makinoshima_2021}.

During this tsunami event, although the first wave arrived at coastal areas within several minutes after the origin time, the officially reported number of tsunami casualties was only two.
The results of this study indicate that one cause of this significantly reduced human loss was early evacuation departure; departure occurred much more quickly than in past tsunami events, such as the 2011 Tohoku tsunami~\cite{makinoshima_2024}.
While people required several tens of minutes to begin tsunami evacuation behaviour in the 2011 Tohoku event, the geolocation data suggested that in the 2024 Noto event, people began evacuation several minutes after the earthquake origin time and reached safe areas within a short period.
The timing of the earthquake occurrence and the Japanese cultural background---i.e., that people typically return to their hometown and spend time with their family members during New Year holidays---are considered the primary causes of this fast evacuation departure trend.
A detailed evacuation survey for the 2011 Tohoku event~\cite{makinoshima_2021} revealed that confirming the safety of family members and family unification were the primary reasons for complex evacuation processes and resulting delayed evacuation departures.
Since the 2011 Tohoku earthquake occurred during the daytime (14:46 JST) on a weekday, many people were engaged in their jobs, which likely separated family members and necessitated additional trips for family unification.
In contrast, the 2024 Noto event occurred in the evening (16:10 JST) during the New Year holiday, when people were typically staying with their family members.
This specific situation helped minimise the time needed for safety confirmation and family unification, leading to the observed fast evacuation departure tendencies.

The geolocation and earthquake observation data suggested that strong ground motion with a maximum peak ground acceleration of 2828 gal, which is comparable to the peak value in the 2011 Tohoku event (2933 gal), was a major trigger for active tsunami evacuation behaviour.
Previous tsunami evacuation surveys revealed that with indigenous knowledge linking earthquakes to potential tsunamis, strong ground motion can be a major trigger for tsunami evacuation~\cite{dudley_2011,lindell_2015,blake_2018,harnantyari_2020,makinoshima_2020}.
However, since these surveys were qualitative, the relationship between the intensity of ground motion and evacuation has not been explored quantitatively.
The geolocation data in this study quantitatively revealed that evacuation activities are related to the intensity of ground motion.
The decay of the ground acceleration observed in this event can be explained well by the widely used empirical equation~\cite{morikawa_2013}.
Therefore, if sufficient data regarding quantitative relationships between ground motion and evacuation activities are collected in future studies, it may become possible to estimate an expected evacuation response from earthquake observations, such as the magnitude and epicentre.
As observed in the case of tsunami earthquakes~\cite{kanamori_1972}, tsunami heights are not always proportional to the perceived seismic intensity.
Further data-driven analyses of tsunami evacuation behaviours may enable a novel disaster response that sends evacuation messages on the basis of predicted evacuation responses.

In addition to the emergency evacuation behaviours, the massive geolocation data revealed long-term behaviours.
The results of this study suggested that after completing tsunami evacuation, people returned to coastal areas long before the tsunami warning or advisory was cancelled.
A previous study of the 2010 Chile tsunami found similar returning behaviour through questionnaires~\cite{kanai_2011}.
However, the actual returning behaviour is generally difficult to clarify with questionnaires since people tend to provide socially desirable responses; i.e., there is a social desirability bias~\cite{grimm_2010}.
For example, an existing study that surveyed both planned and actual tsunami evacuation modes reported that there is a large gap between planning and actual behaviour~\cite{sun_2017}.
Tsunami evacuation surveys based on geolocation data can avoid these psychological biases and reveal real behaviours.
The early return trend revealed by this research provides solid data for improving tsunami warning operations.

While massive geolocation data would be a powerful tool to study evacuation behaviours, such data are likely to be biased.
For example, the samples used in this study have a greater proportion of data on females and a lower proportion of data on the elderly population (Supplementary Table S1), reflecting the demographics of smartphone ownership and the characteristics of the smartphone app.
Additionally, since the detail geolocation data near the user's estimated place of residence were not available for privacy protection (see Methods for details), they were omitted from the analyses in this study.
Therefore, the analysed geolocation data should be considered to have been collected from visitors.
However, for this event, these data are considered a good representation of typical evacuation behaviours in the study areas; this is because, people typically return to their hometown and spend time with their family members during New Year holidays in Japan and because people typically begin evacuation after family unification~\cite{lindell_2015,fraser_2016,blake_2018,makinoshima_2020,makinoshima_2021}.
Given the special circumstances of this tsunami event and typical human behaviour in past tsunami disasters, for this particular case, the analysed geolocation data should provide a good approximation of actual tsunami evacuation dynamics in the study areas.
Upon sufficient understanding of typical human behaviour during tsunamis and careful consideration of sampling biases, analyses of massive geolocation data can provide an overall picture of actual evacuation, which cannot be fully revealed by conventional survey approaches.

\section*{Methods}
\subsection*{Data}
The geolocation data used in this study were collected from users of a move-to-earn smartphone app operated by GeoTechnologies, Inc.
Data are collected and used only with users' consent in this smartphone app.
Additionally, the data are anonymised and privacy-protected so that individuals cannot be identified from the geolocation data.
In particular, location data near the user's estimated place of residence are aggregated into a 125~m mesh for privacy protection.
The geolocation records have a maximum sampling interval of 1~min.
The sample's demographics were provided in Supplementary Table~\ref{tab:s1}.

In addition to the geolocation data, we used geographical data provided by the Geospatial Information Authority in Japan (GSI).
Among the Fundamental Geographical Data provided by the GSI~\cite{gsi}, we used the vector data of coastlines to calculate the shortest distances from coastlines.
The most recent data from before the event were used for the analyses.

To investigate the intensity of ground shaking, we used the seismic data observed by the strong-motion seismograph networks K-NET and KiK-net~\cite{nied}, which are operated by the National Research Institute for Earth Science and Disaster Resilience (NIED).
The peak ground accelerations were obtained by the vector sum of three components of the observation, i.e., east-west, north-south, and up-down components.

\subsection*{Data processing}
Since our focus in this study was on tsunami evacuation behaviours, we analysed the data within the meshes of coastal areas in compliance with the standard grid mesh code (JIS X 0410).
The data within the following meshes were used in this study: Wajima (563607); Suzu (563701, 563712, 563721, 563722); Noto (553770, 553771); Nanao (553647); Himi (553627); Toyama (553701, 553702, 553711, 553712); Itoigawa (553746, 553747); Joetsu (553850, 553851, 553860, 553861, 553862, 553872, 553873); Kashiwazaki (563804); and Niigata (563846, 563856, 563857, 563866, 563867, 563960, 563970, 563971).
Each mesh had an area of approximately 100~km$^2$.
Because the data aggregated for privacy protection could not be used for analysing the evacuation dynamics, we removed such records in the analyses in this study.

Because the continuous recording of geolocation data quickly drains the battery of a smartphone, data recording starts when movements are detected.
This feature of the observations results in a sudden increase in data points in coastal areas during tsunami evacuations.
Therefore, simply taking the average values of data points as snapshots causes some artefacts that do not reflect actual movements.
To address this problem, we used the difference between adjacent data points to obtain the actual movements.
In this study, we first calculated the shortest distance from each data point to the nearest coastline with the 'Join attributes by nearest' function in the QGIS software~\cite{qgis} and added this attribute to the data point.
We then took the difference in the distance to the coastline $\Delta D$ between adjacent data points to determine both landward and seaward movements; i.e., a positive $\Delta D$ indicates landward movement, and negative indicates seaward movement.
If the difference in time $\Delta t$ between adjacent data points is large, it is unclear whether movement continued consistently during $\Delta t$; thus, we considered the difference in distances when the difference in time was less than 5 minutes.
We then organised the movement data ($t + \frac{\Delta t}{2}$, $\Delta D$) into 1~min bins to show the movement trends.

To visualise the geolocation data on a map, we used the contextily package~\cite{contextily} with basemaps provided by OpenStreetMap~\cite{osm} and CartoDB~\cite{carto}.

\section*{Data availability}
The geolocation data used in this study are commercially available from Geo Technologies, Inc., and the authors do not have permission to distribute the data.
The seismic data used in this study were obtained from the strong-motion seismograph networks K-NET and KiK-net, operated by NIED, and are publicly available~\cite{nied} (\url{https://www.doi.org/10.17598/NIED.0004}).
The vector data of coastlines provided by GSI are publicly available~\cite{gsi} (\url{https://www.gsi.go.jp/kiban/}).

\bibliographystyle{unsrt}

\section*{Acknowledgements}
We used seismic observation data provided by the strong-motion seismograph networks K-NET and KiK-net, which are operated by NIED~\cite{nied}.
For map visualisations, we used the contextily Python package~\cite{contextily} with basemaps provided by OpenStreetMap~\cite{osm} and CartoDB~\cite{carto}.
This work was partly supported by the Disaster Resilience Co-creation Center, IRIDeS, Tohoku University.

\section*{Supplementary Information}
\setcounter{figure}{0}
\renewcommand{\thefigure}{S\arabic{figure}}
\renewcommand{\thetable}{S\arabic{table}}

\begin{table}[h]
    \centering
    \caption{Demographics of the geolocation data. The number of samples from subjects who stayed in these cities for more than five minutes 
 on January 1, 2024 are presented. The proportion of each set of samples to the total number of samples in the cities is shown in brackets.}
    \begin{tabular}{cccccccc}
    \toprule
     & & \multicolumn{6}{c}{Age} \\
        &   & 12--19 & 20--29 & 30--39 & 40--49 & 50--59 & 60 < \\ \midrule
        \multirow{10}{*}{Male} & Wajima & 17 (5.1) & 25 (7.5) & 20 (6.0) & 27 (8.1) & 23 (6.9) & 12 (3.6) \\
        & Suzu & 4 (1.5) & 22 (8.4) & 13 (5.0) & 23 (8.8) & 22 (8.4) & 9 (3.4) \\
        & Noto & 16 (6.9) & 20 (8.6) & 13 (5.6) & 21 (9.1) & 19 (8.2) & 7 (3.0) \\
        & Nanao & 36 (4.0) & 62 (6.9) & 58 (6.4) & 82 (9.1) & 79 (8.8) & 27 (3.0) \\
        & Himi & 28 (4.0) & 49 (7.0) & 48 (6.9) & 65 (9.3) & 63 (9.0) & 24 (3.4) \\
        & Toyama & 222 (3.6) & 440 (7.0) & 438 (7.0) & 575 (9.2) & 531 (8.5) & 185 (3.0) \\
        & Itoigawa & 27 (4.9) & 45 (8.2) & 51 (9.3) & 51 (9.3) & 38 (6.9) & 20 (3.6) \\
        & Joetsu & 134 (5.2) & 172 (6.6) & 208 (8.0) & 243 (9.4) & 206 (8.0) & 65 (2.5) \\
        & Kashiwazaki & 50 (4.7) & 68 (6.3) & 87 (8.1) & 100 (9.3) & 99 (9.2) & 53 (4.9) \\
        & Niigata & 457 (4.4) & 761 (7.3) & 794 (7.6) & 939 (9.0) & 874 (8.4) & 313 (3.0) \\ \midrule
        \multirow{10}{*}{Female} & Wajima & 26 (7.8) & 39 (11.6) & 52 (15.5) & 48 (14.3) & 38 (11.3) & 8 (2.4) \\
        & Suzu & 26 (10.0) & 38 (14.6) & 31 (11.9) & 41 (15.7) & 26 (10.0) & 6 (2.3) \\
        & Noto & 20 (8.6) & 28 (12.1) & 32 (13.8) & 31 (13.4) & 18 (7.8) & 7 (3.0) \\
        & Nanao & 62 (6.9) & 125 (13.9) & 128 (14.2) & 128 (14.2) & 89 (9.9) & 25 (2.8) \\
        & Himi & 53 (7.6) & 100 (14.3) & 81 (11.6) & 81 (11.6) & 85 (12.1) & 23 (3.3) \\
        & Toyama & 460 (7.4) & 987 (15.8) & 839 (13.4) & 813 (13.0) & 546 (8.7) & 208 (3.3) \\
        & Itoigawa & 35 (6.4) & 73 (13.3) & 83 (15.1) & 61 (11.1) & 45 (8.2) & 19 (3.5) \\
        & Joetsu & 184 (7.1) & 372 (14.4) & 382 (14.8) & 360 (13.9) & 202 (7.8) & 61 (2.4) \\
        & Kashiwazaki & 67 (6.2) & 161 (15.0) & 152 (14.1) & 142 (13.2) & 75 (7.0) & 21 (2.0) \\
        & Niigata & 730 (7.0) & 1537 (14.7) & 1471 (14.1) & 1395 (13.4) & 903 (8.7) & 250 (2.4) \\
        \bottomrule
    \end{tabular}
    \label{tab:s1}
\end{table}

\end{document}